\documentclass[intlimits,twoside,a4paper]{article}

\usepackage[cp1251]{inputenc}

\usepackage[eqsecnum]{cmpj3}
\issue{2026}{29}{2}{23701}
\doinumber{10.5488/CMP.29.23701}

\title[Halide substitution effects on the photovoltaic properties of Ca$_3$PX$_3$]%
{Halide substitution effects on the photovoltaic properties of Ca$_3$PX$_3$ (X = F, Cl, Br, I) perovskites: advancing solar cell efficiency%
 }
\author[P. Dhariwal et al.]{P. Dhariwal\orcid{0009-0005-3665-4342}\refaddr{label1},
       D. Prakash\orcid{0000-0003-3729-683X}\refaddr{label2}, 
       K. D. Verma\orcid{0000-0002-5492-2997}\refaddr{label1}\thanks{Corresponding author: \email{kdverma1215868@gmail.com}.}, 
       A. Kumari\orcid{0009-0003-0062-1093}\refaddr{label1}, 
       P. K. Kamlesh\orcid{0000-0001-7361-3519}\refaddr{label3}, 
       A.~S.~Verma\orcid{0000-0001-8223-7658}\refaddr{label4,label5}\thanks{Email: \email{ajay\_phy@rediffmail.com}.}}
\addresses{
\addr{label1} Materials Science Research Laboratory, Department of Physics, Shri Varshney College, Raja Mahendra Pratap Singh University, Aligarh, 202001 Uttar Pradesh, India
\addr{label2} School of Computer Science and Engineering, Faculty of Engineering, Shri Mata Vaishno Devi University, Kakryal, Katra, 182320 J\&K, India
\addr{label3} Department of Physics, Poornima University, Jaipur, 303905 Rajasthan, India
\addr{label4} Department of Allied Sciences, Graphic Era Deemed to be University, Dehradun, 248002 Uttarakhand, India
\addr{label5} University Centre for Research \& Development, Department of Physics, Chandigarh University, 140413 Mohali, Punjab, India
}
\date{Received 23 August 2025; revised 11 December 2025; accepted 28 January 2026; published 29 June 2026}

\begin{document}

\maketitle

\begin{abstract}
Herein, the fundamental physical characteristics like structural, electronic, optical parameters of the Ca$_3$PX$_3$ (X = F, Cl, Br, I) materials have been investigated for their potential optoelectronic applications, parti\-cu\-larly for solar cells and related devices. To the crystallographic investigations, Ca$_3$PI$_3$ has the most stable configuration among all investigated materials. From the band structure analyses of these materials indicate that all materials have a direct bandgap in the range of 2.0~eV to 3.788~eV, which makes them ideal for light absorption. For the photovoltaic applications, we have analysed first-principles spectroscopic screening limited maximum efficiency~(SLME) which  confirms that the Ca$_3$PI$_3$ material exhibits the highest solar cell efficiency 29.6\% and Ca$_3$PF$_3$ and shows lower efficiency for solar cell suitability 0.6\%.  Thus, these results demonstrate the real potential and abilities of halide substitution to tune the materials for particular optoelectronic devices.
\end{abstract}

\section{Introduction}

%\doclicenseThis
Due to their remarkable optoelectronic, mechanical and structural properties, perovskite materials are gaining a lot of attention in renewables research~\cite{ref1,ref2,ref3}. These materials are distinguished by their ABX$_3$ crystal structure, in which halide ions construct a cubic skeleton and cations situated at the corners make these ionic materials fantastic candidates for solar energy conversion and next-generation technologies~\cite{ref4,ref5,ref6}. Early studies of calcium-based materials have revealed good promise in solar cells, quantum devices, and optoelectronics~\cite{ref7,ref8}. They also exhibit good bandgaps, high light absorption, and mechanical performance making them good candidates for advanced photovoltaic (PV) and energy storage devices. Despite the dominance of silicon in the current PV market, its high manufacturing cost and non-ideal bandgap hinder a further progress~\cite{ref9,ref10}. An exceptional development trajectory has been made, with the power conversion efficiency (PCE) improving from 3.8\% to greater than 25\% in less than ten years, stimulating considerable interest in hybrid (organic-inorganic) perovskite solar cells~\cite{ref11}. Organic-inorganic lead halide perovskites present very good efficiencies but do indeed suffer from the aspects of toxicity, device hysteresis, and thermal instability due to organic cations. As a result, researchers are increasingly focusing on inorganic, Pb-free alternatives such as Ca$_3$PX$_3$ materials, where X = F, Cl, Br, I. These substances offer a great potential to address the key challenges related to stability and environmental safety while maintaining efficient light-harvesting capabilities. 
These materials belong to the A$_3$BX$_3$ family, where A represents $\mathrm{Mg}^{2+}$, $\mathrm{Ca}^{2+}$, $\mathrm{Sr}^{2+}$, or $\mathrm{Ba}^{2+}$, X is $\mathrm{F}^{-}$, $\mathrm{Cl}^{-}$, $\mathrm{Br}^{-}$, or $\mathrm{I}^{-}$, and B can be $\mathrm{N}^{3-}$, $\mathrm{P}^{3-}$, $\mathrm{As}^{3-}$, or $\mathrm{Sb}^{3-}$~\cite{ref12,ref13}. This group of materials shares a cubic structure with the Pm$\bar{3}$m space group, resembling the structure of traditional halide perovskites, yet they often exhibit different coordination geometries and bonding situations which give rise to unique physical properties. The variation of halogen atoms systematically modifies ionic size and electronegativity which leads to substantial changes in the lattice constants, bonding environment, and ultimately the electronic and optical characteristics of the substances being synthesized. The potential to work with light, earth-abundant elements such as calcium and phosphorus also makes Ca$_3$PX$_3$ systems highly desirable for scalable and sustainable applications.

Notable for its high transparency, efficient light emission, and adjustable bandgap, Ca$_3$PBr$_3$ is ideally suited toward optoelectronic devices such as light-emitting diodes (LEDs), photo-detectors and other photonic technologies~\cite{ref14}. The high transparency, adjustable bandgap, based on stoichiometry and purity, and efficiency of Ca$_3$PBr$_3$ put it in a strong position to advance PV and photonic technologies like photo-detectors, solar cells, and LEDs, with its tuneable bandgap, stability and absorption efficiencies~\cite{ref15}. Recent first-principles density functional theory (DFT) studies using pseudo-potentials within the Quantum Espresso simulation code have been performed to determine the configurational, optical and electronic properties of Ca$_3$PX$_3$ (X = I, Br, Cl) under the action of strain and spin-orbit coupling~(SOC) in the band structure. These studies indicate that strain engineer alters the bandgap qualitatively and that can enhance the performance under various mechanical conditions~\cite{ref16}. In a study, Shimul et al.~\cite{ref17} have highlighted the promising optoelectronic potential of lead-free halide perovskites, A$_3$PCl$_3$ (A = Mg, Ca), under pressure. Their tuneable bandgap and high PCE make them as new and eco-friendly candidates for next-generation solar applications. Ghosh et al.~\cite{ref18} have conducted DFT based investigations on A$_3$BX$_3$ (A = Ca/Sr, B = P/As, X = I/Br) halide perovskites. They observed that these materials exhibit tuneable bandgaps and good light absorption in the optical range with low synthesized energy. Notably, the solar cells made of Ca$_3$PBr$_3$ and the solar cell based on Ca$_3$PI$_3$ achieved a PCE of 13.66\% and 24.7\%,  respectively. They highlighted the utility of these materials to next-generation PV applications.

A prior study has described the mechanical, optical, and electronic characteristics of Ca$_3$PCl$_3$ using DFT simulations. This study reported that the absorption behaviour of Ca$_3$PCl$_3$ depends on the strain of the material. It characterized that under different strain conditions the optical absorption peaks of Ca$_3$PI$_3$ shift from the UV region to the regions of visible light~\cite{ref19}. Ca$_3$PI$_3$ has a very symmetric structure, with a Pm$\bar{3}$m (\#221) space group like other cubic halide perovskites~\cite{ref20}. This symmetry is important for the transitions at the band edge states, reinforcing their potential for use in PV applications~\cite{ref21}. Another previous work focused on how SOC and strain confer the deviations in the bandgap and electronic characteristics of Ca$_3$PI$_3$. It focused on the optical absorption behavior, in which red and blue shifts of the dielectric peaks were observed as a function of strain. The optoelectronic behavior of Ca$_3$PI$_3$ can be tuned under biaxial strains, which shows its potential in solar cells and other optoelectronic applications. Other studies of A$_3$BX$_3$ class of materials also provide similar understanding of the important fundamental optical, electronic, mechanical, and structural properties of these substances~\cite{ref22}.

The above literature review shows that in Ca$_3$PX$_3$ compounds, where X = F, Cl, Br, and I, a few of them have been explored using pseudo-potential while the  remaining ones are unexplored. Therefore, using full potential DFT (FP-DFT), a detailed investigation of their crystallographic, electronic, and optical behaviour is essential to unlock their maximum efficiency and guide future experimental and device-level advancements. Therefore, the present work aims at  determining the structural, optical, and electronic characteristics of Ca$_3$AX$_3$ materials, where X = F, Cl, Br, I, through FP-DFT methods implemented in WIEN2K simulation package and compute their spectroscopic limited maximum efficiency (SLME). By systematically analyzing the effects of anionic and cationic substitutions, this study seeks to identify the non-toxic inorganic candidates with high PV and optoelectronic potential, thereby supporting the advancement of sustainable energy technologies.

\newpage
\section{Methodology}

The fundamental physical characteristics including structural, optical, and electronic characteristics of the Ca$_3$PX$_3$ (X = F, Cl, Br, I) halide materials, were thoroughly examined utilizing the all-electron ``Full-Potential Linearized augmented plane wave + local orbitals'' (FP-LAPW + lo) techniques, as incorporated in the WIEN2k simulation package~\cite{ref23}. Geometry optimization and ground-state property evaluations were conducted using the ``Generalized gradient approximation'' formulated by Perdew--Burke--Ernzerhof (PBE-GGA)~\cite{ref24}. To minimize the overlap and maximize the sphere filling, appropriate muffin-tin radii ($R_{\text{MT}}$) were chosen, and the smallest $R_{\text{MT}}$ multiplied by the plane-wave cut-off parameter was set to $R_{\text{MT}}K_{\text{max}} = 7.5$. A Monkhorst-Pack grid of $10\times10\times10$ $k$-points was employed for determining the structural optimization and electronic structure. The angular momentum expansion within the atomic spheres was set to $l_{\rm max} = 10$, and the maximum modulus for $G$ vectors ($G_{\text{max}}$) was fixed at $12~\mathrm{a.u}^{-1}$. An energy cut-off of $-6.0$~eV was applied to discriminate between the valence and core states to ensure no charge leakage. Self-consistent field (SCF) cycles were iterated until convergence criteria were met, with thresholds of $0.001e$ for total charge density and 0.0001~Ry for total energy variation. Structural relaxations continued until the equilibrium lattice parameters minimized the total energy, ensuring a stable configuration. The optimized structures were then used for calculations of total and partial density of states (DOS), electronic band structures, and optical parameters, such as real and imaginary parts of the absorption coefficient, refractive index, dielectric function, energy loss function, etc. In order to obtain a much more  accurate estimation of the bandgap and improve the exchange-correlation potential, the Tran--Blaha modified Becke--Johnson (TB-mBJ)~\cite{ref25} approach was combined with PBE-GGA. To capture the fine features in the dielectric function and other optical spectra, a denser grid of $20\times20\times20$ $k$-points was employed for the optical property calculations. The SLME is computed using SLME code~\cite{ref26}.

\section{Results and discussion}

\subsection{Structural properties}

\begin{figure}[h]
	\centering
	\includegraphics[scale=0.7]{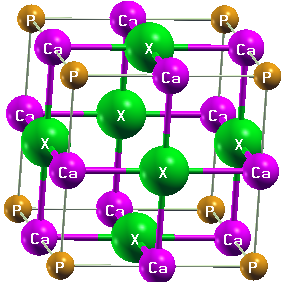}
	\caption{(Colour online) Crystal structure of Ca$_3$PX$_3$ compounds.}
	\label{fig:1}
\end{figure}

\begin{figure}[htb]
	\centering
	\includegraphics[scale=0.55]{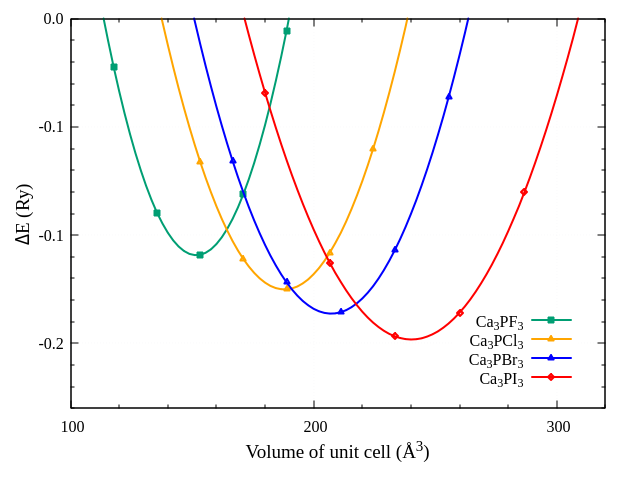}
	\caption{(Colour online) Plots for lattice volume ( \AA$^{3}$) vs the change in ground state energy (Ry).}
	\label{fig:2}
\end{figure}

The Ca$_3$PX$_3$ (X = F, Cl, Br, I) materials hold unique atomic arrangements, forming cubic structures. These structures belong to the Pm$\bar{3}$m (\#221) space group, where Ca, P, and X atoms lie at (0.5, 0.5, 0), (0.5, 0.5, 0.5) and (0.5, 0, 0) Wyckoff sites, correspondingly. The unit cell and primitive cell structures of the materials studied are represented in figure~\ref{fig:1}. 
At equilibrium, the structural characteristics, such as lattice constant ($a$ in \AA), ground-state energy ($E$), bulk modulus at zero pressure ($B$), and its first pressure derivative ($B'$) were calculated using the Birch--Murnaghan's equation of state~\cite{ref27,ref28}. Figure~\ref{fig:2} displays the correlation between the lattice volume (in \AA$^3$) and the energy change ($\Delta E$ in Ry) for four Ca$_3$PX$_3$ compounds: Ca$_3$PF$_3$, Ca$_3$PCl$_3$, Ca$_3$PBr$_3$, and Ca$_3$PI$_3$. The downward slope of each curve shows that compressing the lattice leads to lower energy, as atoms are brought closer together and optimize the bonding interactions. However, when compression exceeds a certain limit (post-equilibrium), repulsive interactions between atoms dominate, causing the energy to rise. The unit cell volume corresponding to the minimum energy of a material represents its most stable structure, from which the lattice constant at equilibrium can be determined. The structural characteristics for the selected compounds are summarized in table~\ref{tab:1}. The lattice constant of Ca$_3$PX$_3$ ranges from 5.321~\AA~to 6.200~\AA, showing an increase with the growing size of the X-atoms. This trend is consistent with theoretical predictions~\cite{ref16,ref18} and is further validated by the observations in figure~\ref{fig:2}.

Moreover, the susceptibility of the materials to compression is evaluated using the bulk modulus, with a higher value indicating a reduced compressibility. Table~\ref{tab:1}  illustrates that $B$ is inversely proportional to the lattice constants of the materials. Here, the order of compressibility is determined as Ca$_3$PI$_3$> Ca$_3$PBr$_3$> Ca$_3$PCl$_3$> Ca$_3$PF$_3$. An essential parameter, denoted as $B'$, illustrates how the bulk modulus changes with an increasing pressure. Here, the positive $B'$ value indicates that the materials demonstrate an increased resistance to deformation under higher pressure. The ground-state energy ($E$) trend shows that Ca$_3$PI$_3$ is the most stable material, while Ca$_3$PF$_3$ material is the least stable. The observed changes in the lattice constant with varying X-atom sizes further confirm the accuracy and reliability of this study.

\begin{table}[htb]
	\caption{Structural parameters of Ca$_3$PX$_3$ (X = F, Cl, Br, I) materials.}
	\label{tab:1}
	\begin{center}
		\begin{tabular}{|c|c|c|c|c|c|c|c|}
			\hline
			\textbf{Materials} & \textbf{$a$ (\AA)} & \textbf{$B$ (GPa)} & \textbf{$B'$} & \textbf{$E$ (Ry)} & \textbf{{Pressure (GPa)}} & \multicolumn{2}{c|}{\textbf{Bandgap (eV)}} \\
			\hline
			&&&&& & \textbf{PBE} & \textbf{PBE+TB-mBJ} \\
			\hline
			
			Ca$_3$PF$_3$ 
			& 5.321 & 48.811 & 4.008 & $-5366.809$ & {$-0.12$} & 2.518 & 3.788 \\
			\hline
			
			Ca$_3$PCl$_3$ 
			& 5.713 & 39.622 & 4.960 & $-7536.595$ & {$-0.195$} & 1.978 & 2.858 \\
			& 5.725$^{b}$ &  &  &  &  & 2.205$^{b}$ &  \\
			\hline
			
			Ca$_3$PBr$_3$ 
			& 5.904 & 37.146 & 4.189 & $-20407.376$ & {0.054} & 1.819 & 2.553 \\
			& 5.91$^{a}$ &  &  &  &  & 1.957$^{a}$ &  \\
			& 5.923$^{b}$ &  &  &  &  & 1.950$^{b}$ &  \\
			\hline
			
			Ca$_3$PI$_3$ 
			& 6.200 & 32.019 & 4.603 & $-47481.678$ & {$-0.052$} & 1.469 & 2.000 \\
			& 6.21$^{a}$ &  &  &  &  & 1.582$^{a}$ &  \\
			& 6.206$^{b}$ &  &  &  &  & 1.490$^{b}$ &  \\
			\hline
			
			\multicolumn{8}{l}{$^{a}$Reference~\cite{ref18}; \quad $^{b}$Reference~\cite{ref16}} 
		\end{tabular}
	\end{center}
\end{table}

\subsection{Electronic properties}

\begin{figure}[h]
	\centering
	
	% Row 1
	\begin{minipage}{0.35\textwidth}
		\centering
		\includegraphics[width=\linewidth]{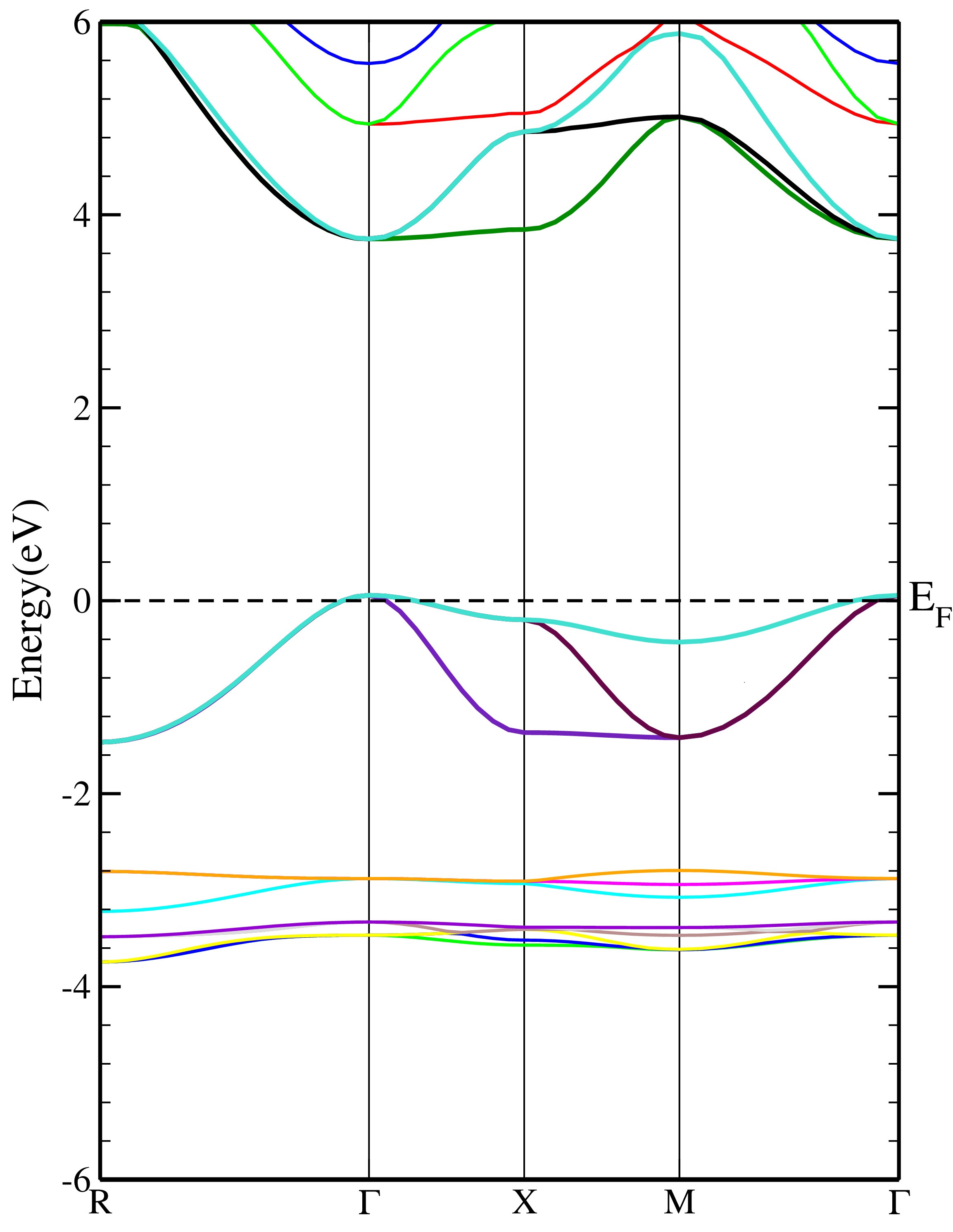}
		\\ (a) Ca$_3$PF$_3$
	\end{minipage}
	\hspace{0.05\textwidth}
	\begin{minipage}{0.35\textwidth}
		\centering
		\includegraphics[width=\linewidth]{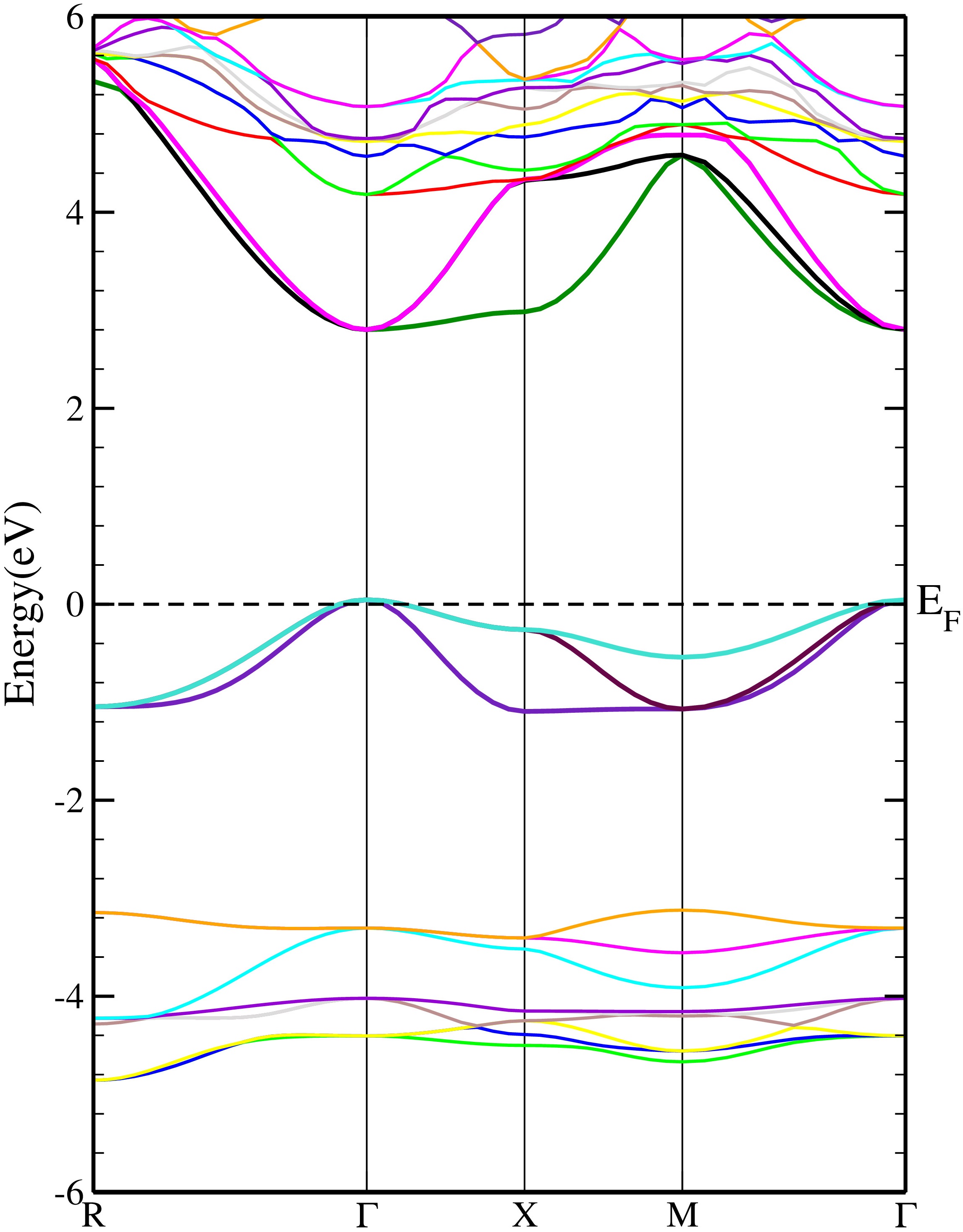}
		\\ (b) Ca$_3$PCl$_3$
	\end{minipage}
	
	\vspace{0.3cm} % space between rows
	
	% Row 2
	\begin{minipage}{0.35\textwidth}
		\centering
		\includegraphics[width=\linewidth]{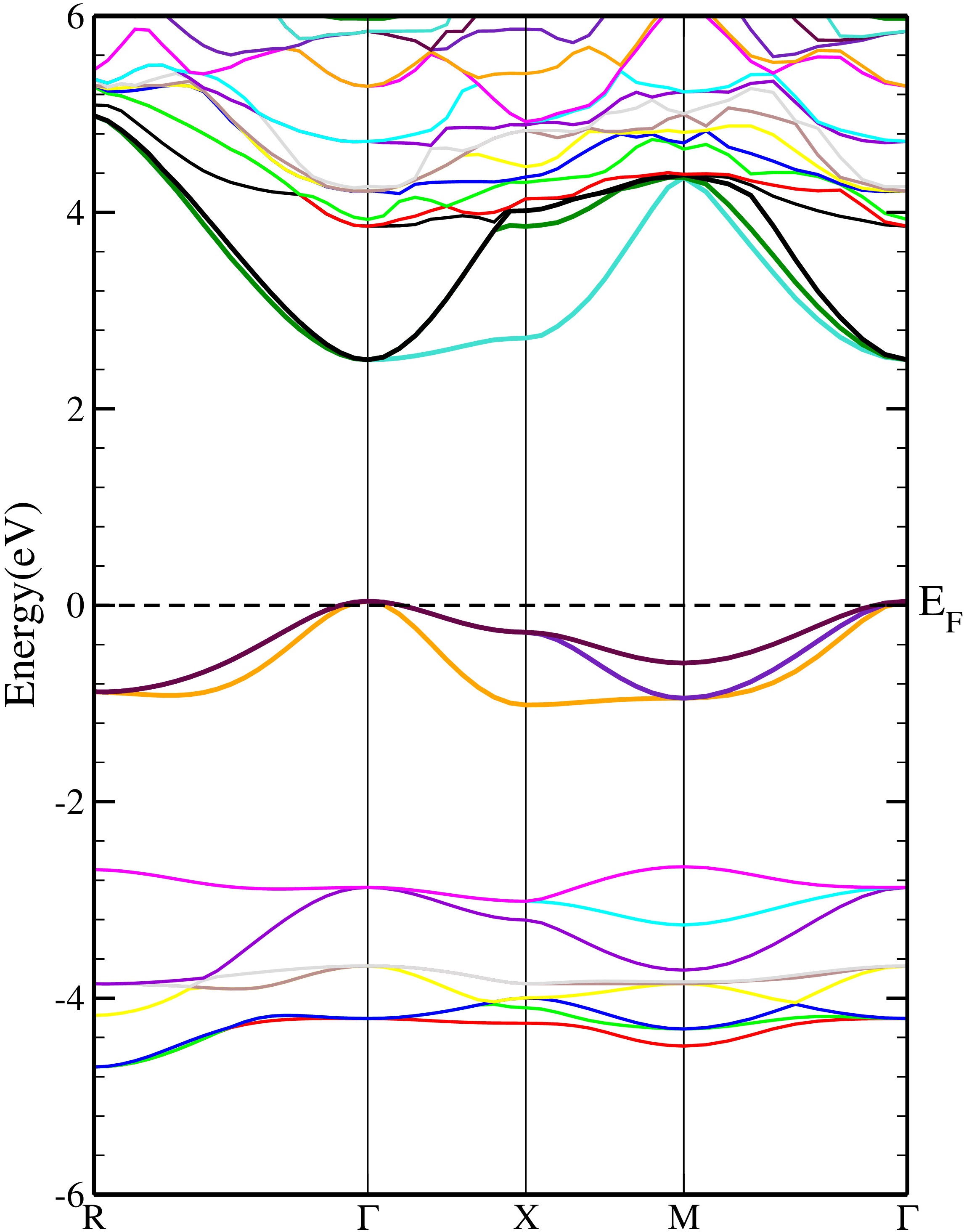}
		\\ (c) Ca$_3$PBr$_3$
	\end{minipage}
	\hspace{0.05\textwidth}
	\begin{minipage}{0.35\textwidth}
		\centering
		\includegraphics[width=\linewidth]{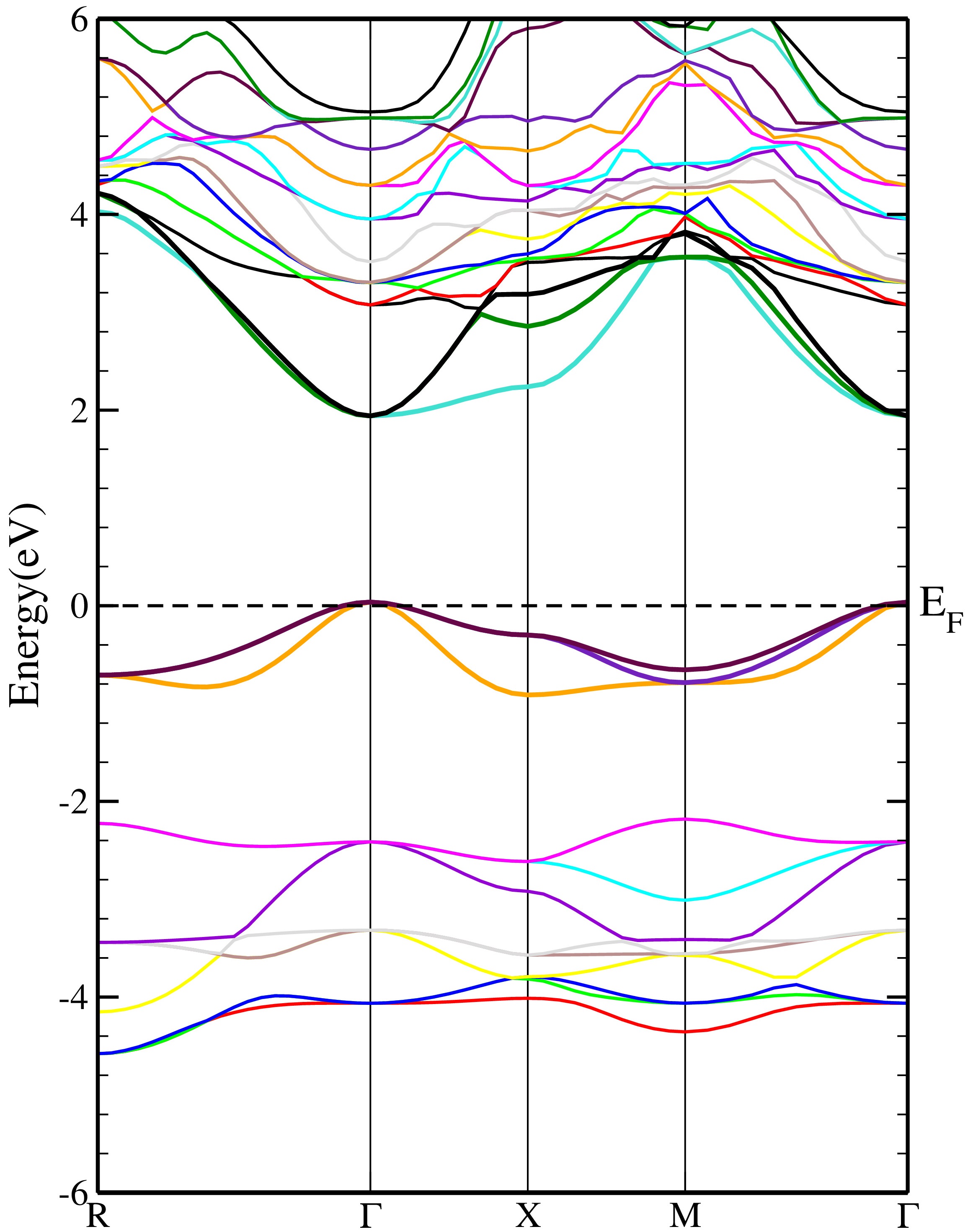}
		\\ (d) Ca$_3$PI$_3$
	\end{minipage}
	
	\caption{(Colour online) Band structure of Ca$_3$PX$_3$ halides obtained using  PBE+TB-mBJ potential.}
	\label{fig:3}
\end{figure}

A thorough analysis of the electronic band structure is done utilizing both the PBE and PBE+TB-mBJ method. Table~\ref{tab:1} represents the bandgap values of the investigated compounds utilizing PBE and PBE+TB-mBJ approaches. The bandgap of these materials calculated using PBE and PBE+TB-mBJ approaches yields in the ranges of 1.469--2.518 eV and 2.000--3.788 eV, respectively. The band structure plots (figure~\ref{fig:3}) of the Ca$_3$PX$_3$ compounds show the key insights into their electronic properties and suitability for solar cell devices. The presence of direct bandgaps at the $\Gamma$-point in these compounds suggests that these are ideal for efficient light absorption and electron excitation, making them excellent candidates for high-performance solar cells.

The moderate bandgaps in these materials allow for effective absorption of sunlight, with Ca$_3$PI$_3$ showing a particularly promising bandgap for the use in perovskite-based solar cells, known for their high efficiencies. Near the Fermi level $E_{f}$, at $\Gamma$-point, the band structure curves for different states appear to coincide or have similar energy levels, suggesting a degenerate state. This degeneracy enhances the availability of the energy states for electron movement, which in turn boosts the electrical conductivity. For solar cell applications, this property is essential since it facilitates the efficient movement of photoexcited electrons from the valence to conduction band, a key process of converting light into electrical energy.

\begin{figure}[h]
	\centerline{\includegraphics[scale=0.6]{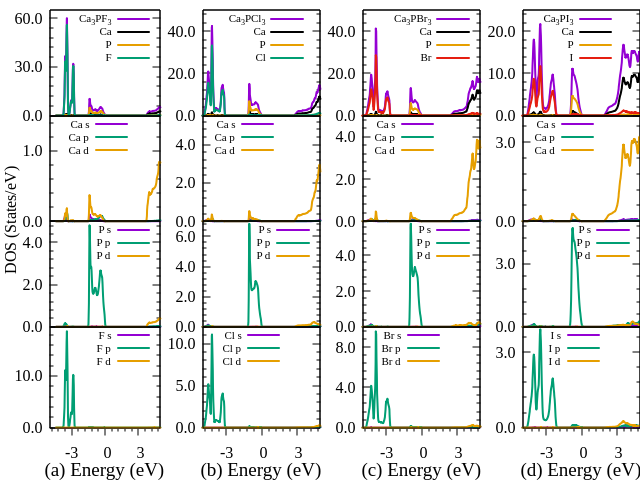}}
	\caption{(Colour online) PDOS plots of Ca$_3$PX$_3$ (X = F, Cl, Br, I) materials.} \label{fig:4}
\end{figure}

To gain a more detailed insight into the band structure, we computed the DOS for these materials. The partial and total DOS at the equilibrium lattice constant were calculated utilizing PBE+TB-mBJ approach, which are depicted in figure~\ref{fig:4}. A clear correlation is observed between the band structures with the partial and total DOS plots. The total DOS can be separated into two distinct regions: one corresponding to the conduction band and the other one to the valence band. Despite Ca$^{2+}$ ion being nominally a $d$-empty ion, the apparent $d$-character shown in the DOS emerges from hybridization between Ca $3d$ and P $3p$ states in the crystal field enviroment. This hybridization effect redistributes electronic density and generates $d$-projected contributions in the valence and conduction region. Analogous hybridization interactions also give rise to the small $d$-state features observed for the halide atoms (X = F, Cl, Br, I), which are interactions with the Ca and P and not necessarily indicative of a true occupied $d$-state. 

For all studied materials, the primary contribution near the top of the valence band of the materials is from the phosphorus atom (P) and predominantly through its $p$ orbital, making them important for both electronic and optical interactions. In the valence band, the Ca-$p$ states dominate at the lower energy ranges, while the Ca-$d$ states govern at slightly higher energy ranges. In their conduction band, Ca-$d$ states dominate near the conduction band edge. The contribution of the X atoms near the minima of the conduction band is found to be negligible, although, a small contribution of X-$d$ states is observed, which increases as the size of the X-atom increases. Thus, in these materials, the Ca and P orbitals dominate the electronic states, although, the contribution of halide orbitals extend across a broader energy range, influencing the electronic properties of the material. This contribution becomes more pronounced as the size of the halide atom increases. Ca$_3$PI$_3$ shows the strongest contribution of iodine, particularly in the orbitals $d$ and $p$ near the Fermi level, making it an excellent candidate for optical absorption in the visible range, ideal for PV devices. Therefore, since we can tune these contributions via changes in the halogen atoms, we can optimize the properties of the material for optoelectronic devices like solar cells and LED lights.
%\newpage
%%%%%%%%%%%%%%%%%%%%%Optical properties%%%%%%%%%%%%%%%%%%%%%%

\subsection{Optical properties}

The optical behaviour of a substance indicate its conduct with electromagnetic (EM) radiation. Optical characteristics are typically designated by the complex dielectric function, which describes how electrons respond to photons in a compound. For small wave vectors, the complex dielectric function provides information about absorption and scattering of light. The dielectric function is described as~\cite{ref29}: 
\begin{align}\label{eqn:dielectric}
	\varepsilon(\omega) &= \varepsilon_{1}(\omega) + \ri\varepsilon_{2}(\omega).
\end{align}
Here, $\omega$ represents the angular frequency of the incident EM radiation. The real part of the dielectric function, $\varepsilon_{1}(\omega)$, is related to the  capability of the material to store energy through electronic polarization and its refractive index variation with light frequency, known as anomalous dispersion. The imaginary part,  $\varepsilon_{2}(\omega)$, defines the optical absorption, representing the capability of a substance to absorb incident EM radiation. The imaginary component of the dielectric function is given by~\cite{ref30}:
%%%%%%%%%%%%%%%%%%%%%%%equation%%%%%%%%%%%%%%%%%%%%%%
\begin{align}\label{eqn:im_dielectric}
	\varepsilon_{2}(\omega) &=\frac{4e^2\piup^2}{\omega^2 m^2}\sum_{i,j}\int\big{|}\big{<}i|M|j\big{>}\big{|}^2f_i(1-f_i)X\delta(E_f-E_i-\hbar\omega)\, \rd^3k.
\end{align}
%%%%%%%%%%%%%%%%%%%%%%%%%%%%%%%%%%%%%%%%%%%%%%%%%%%%%%%%%%
In the above equation, $M$ denotes the dipole matrix element, which quantifies the interaction between the initial and final states during the absorption process. $i$ and $j$ represent the initial and final states (valence and conduction bands, respectively). $f_i$ is the Fermi distribution function, which states the probability that an electron occupies a state in the valence band at a given energy. $\delta(E_f-E_i-\hbar\omega)$ describes the energy conservation condition during the photon absorption process. It ensures that the energy difference between the conduction and valence bands at a specific $k$-point matches the photon energy $\hbar$$\omega$. $e$ is the electronic charge, $\hbar$ is the reduced Planck's constant, $\omega$ is the angular frequency of the incident photon, and $m$ is the electron mass. Using the Kramers--Kroning relation~\cite{ref31}, the $\varepsilon_{1}(\omega)$ can be computed as:
%%%%%%%%%%%%%%%%%%%%%%%%%%%%%equation%%%%%%%%%%%%%%%%%%%%%%%%%
\begin{align}\label{eqn:real_dielectric}
\varepsilon_{1}(\omega) = 1 + \frac{2}{\piup}P\int_{0}^{\infty}\frac{\omega'\varepsilon_{2}(\omega')}{\omega'^2-\omega^2}\,\rd\omega'.
\end{align}
%%%%%%%%%%%%%%%%%%%%%%%%%%%%%%%%%%%%%%%%%%%%%%%%%%%%%%%%%%%%%
Here, $P$ denotes the principal value of the integral, which accounts for the singularity at $\omega' = \omega$. $\varepsilon_{2}(\omega')$ is the imaginary part of the dielectric function at a different frequency $\omega'$, and it is used to compute the real part $\varepsilon_{1}(\omega)$. By analysing both the real and imaginary components of the dielectric function, $\varepsilon_{1}(\omega)$ and $\varepsilon_{2}(\omega)$, other optical properties such as optical conductivity $\sigma(\omega)$, reflectivity $R(\omega)$, extinction coefficient $k(\omega)$, refractive index $n(\omega)$, absorption coefficient $\alpha(\omega)$, and energy loss function $E_{\rm loss}(\omega)$ can be derived~\cite{ref32}. The optical properties for the equilibrium structure were computed for incident EM radiation energies ranging from 0 to 13~eV and are shown in figures~\ref{fig:5}--\ref{fig:8}.

\subsubsection{Real and imaginary part of dielectric function}

\begin{figure}[htb]
	\centerline{\includegraphics[scale=0.3]{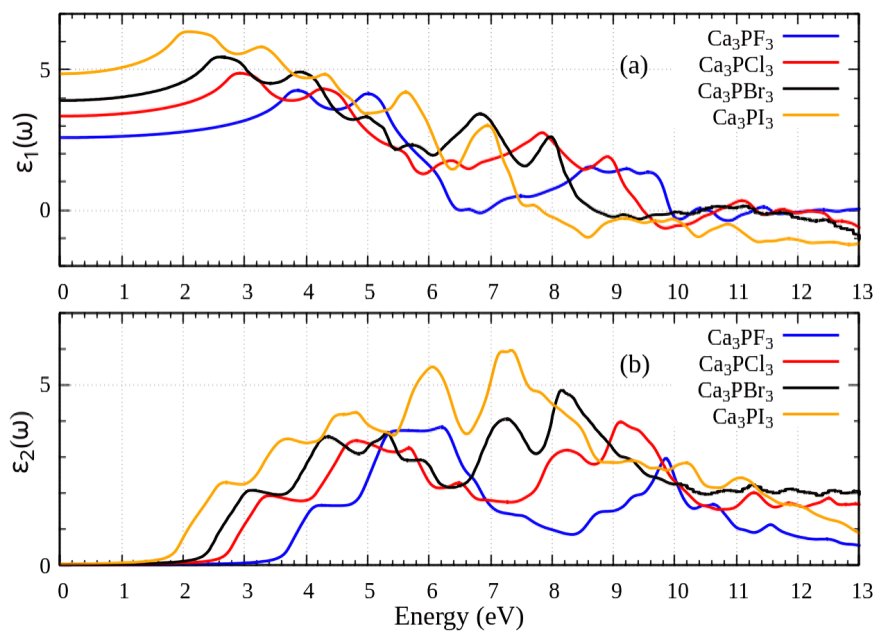}}
	\caption{(Colour online)  Variation in (a) $\varepsilon_{1}$($\omega$) and (b) $\varepsilon_{2}$($\omega$) of dielectric tensor spectra of Ca$_3$PX$_3$ (X~=~F, Cl, Br, I) compounds materials.} \label{fig:5}
\end{figure}

Figures~\ref{fig:5}(a) illustrate how $\varepsilon_{1}$($\omega$) varies with the  changes in the incident EM radiation energy. $\varepsilon_{1}$($\omega$) reflects the capability of a substance to polarize when exposed to an electric field, with peaks in $\varepsilon$ signifying the resonance with EM radiation at specific energies. Its static value, $\varepsilon_{1}$(0), represents the dielectric constant of the material. Figure~\ref{fig:5}(a) demonstrates the values of $\varepsilon_{1}$(0) for Ca$_3$PX$_3$ (X = F, Cl, Br, I) materials which are 2.58, 3.34, 3.90, and 4.85, respectively. This indicates that the dielectric constant of the chosen substances rises as the size of the halogen atom (X) grows. The trend reveals that as the atomic radius of the X-atom rises, the capability of the substance to respond to the incident EM radiation becomes more pronounced. A greater value of $\varepsilon_{1}$(0) suggests that the compound displays a stronger polarization response to the applied electric field of the incoming EM radiation. All four materials show an increase in $\varepsilon_{1}$($\omega$) with the rising energy, and make a maximum in the range 3.85, 2.93, 2.60, and 2.08 eV, respectively, suggesting stronger polarization at these energies. Ca$_3$PI$_3$ exhibits the highest $\varepsilon_{1}$($\omega$) values, indicating a more efficient interaction with light, particularly in the lower energy range. Following the peak, $\varepsilon_{1}$($\omega$) decreases, displaying additional oscillations and ultimately becoming negative at 6.71, 9.48, 8.69, and 7.85 eV, with further peaks and valleys observed at higher energies. This suggest that, at higher energies, the response of the material becomes less efficient, probably due to the interband transitions, electron scattering, or other processes that dampen polarization at these energies. The presence of further peaks and valleys indicates complex interactions within the electronic structure of the material, highlighting the intricacies of its response to high-energy EM radiation. These values of $\varepsilon_{1}$(0) along with previously computed values~\cite{ref16} for some of the considered materials are mentioned in table~\ref{tab:2}. The significant difference in $\varepsilon_{1}$(0) is observed due to the different computational approaches that were implemented. Specifically, the present study employed a FP-LAPW + lo method for all electrons, and combined it with the modified Becke--Johnson (TB-mBJ) method for calculating the self-consistent field cycles. This approach produces a more accurate bandgap than the norm-conserving pseudopotentials used in reference~\cite{ref16} under the PBE-GGA approximation. This difference in methodologies also affects how core states, and the quality of the basis sets, are treated, as well as the exchange and correlation functionals and the k-point densities used in the electronic transitions that contribute to the static dielectric constant.

%%%%%%%%%%%%%%%%%%%%%%%%Fig%%%%%%%%%%%%%%%%%%%%%%%%%%%%%%%%%

The changes in the imaginary part, $\varepsilon_{2}$($\omega$), with the energy is displayed in figures~\ref{fig:5}(b). Ca$_3$PI$_3$ shows intense peaks in $\varepsilon_{2}$($\omega$), indicating that it absorbs light most efficiently in the UV and visible spectra, positioning it as a promising material for solar cell applications. By contrast, Ca$_3$PF$_3$ exhibits a more gradual growth in $\varepsilon_{2}$($\omega$), suggesting a weaker light absorption. It exhibits a broad absorption peak near 4 eV and 5 eV and a secondary feature around 10 eV, signifying multiple electronic transitions with distinct energy thresholds. In Ca$_3$PI$_3$, sharper and more pronounced peaks are observed at lower energies, particularly around 2.6~eV, which indicate the presence of intermediate electronic states or additional excitations. Thus, materials like Ca$_3$PI$_3$, with strong light absorption and efficient polarization, are well-suited for PV devices, while others, like Ca$_3$PF$_3$, may be less efficient in such applications. The variation in dielectric properties with different halide elements (F, Cl, Br, I) allows for tuning these materials for specific uses in solar cells and other optoelectronic technologies.

\subsubsection{Electrical conductivity and absorption coefficient} 

\begin{figure}[htb]
	\centerline{\includegraphics[scale=0.3]{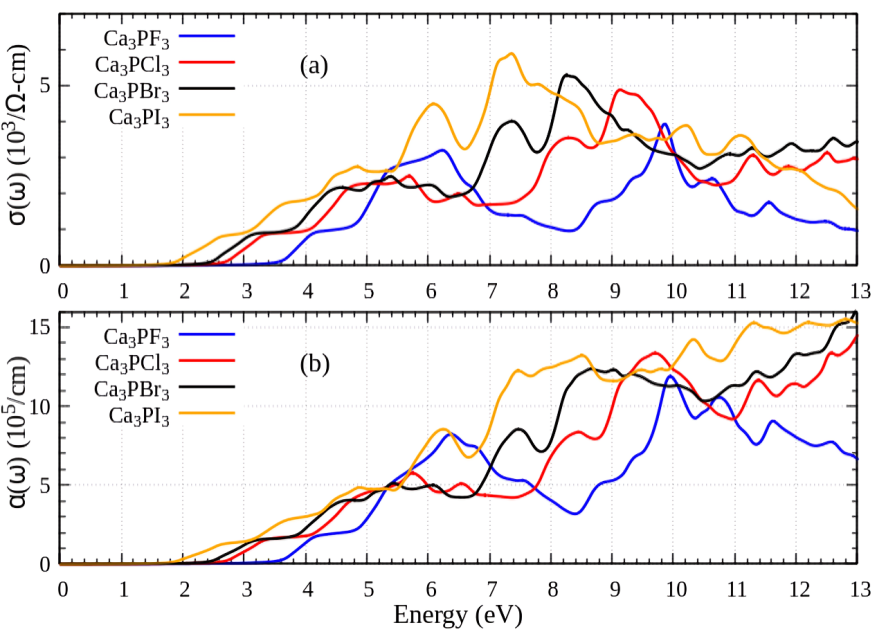}}
	\caption{(Colour online) Variation in (a) $\sigma$($\omega$)  and  (b) $\alpha$($\omega$)  for Ca$_3$PX$_3$ (X = F, Cl, Br, I) compounds.} \label{fig:6}
\end{figure}

Optical conductivity, $\sigma$($\omega$), is a key parameter used to describe the behaviour of photoelectrons as they move under the action of EM radiation. Figure~\ref{fig:6}(a) shows the variation of $\sigma$($\omega$) with energy, spanning from 0 eV to 13 eV. From this figure, it is manifest that all materials show a noticeable increase in $\sigma$($\omega$) as energy rises, which is characteristic of semiconductor materials where the conductivity improves as electrons are excited from the valence band to the conduction band, enabling a  better electron flow under the action of EM radiation. Among the materials, Ca$_3$PI$_3$ exhibits the highest conductivity, especially at lower energies, suggesting a greater capability for charge carriers to flow in response to an electric field. Ca$_3$PBr$_3$ and Ca$_3$PCl$_3$ also show a relatively higher conductivity compared to Ca$_3$PF$_3$, which has the lowest values. This difference indicates that the nature of the halide (F, Cl, Br, I) strongly influences the conductivity, with iodine (in Ca$_3$PI$_3$) contributing to an  enhanced charge transport. The absorption coefficient of a substance indicates the extent to which incident radiation is absorbed by a unit thickness of the substance. The higher is the absorption coefficient, the more effectively the material transfers electrons from the valence band to the conduction band, enhancing its capability to absorb and convert EM radiation into electrical energy. Figure~\ref{fig:6}(b) demonstrates the energy of the incident radiation and its relative absorption coefficient notation $\alpha$($\omega$), for Ca$_3$PX$_3$ (X = F, Cl, Br, I). For the absorption coefficient, shown in the lower panel, all materials show peaks at specific energies. 
Ca$_3$PI$_3$ again stands out with a strong absorption peak, especially around the 2 eV to 7.5 eV range, indicating that it efficiently absorbs light in both the visible and near-UV ranges. This strong absorption capability makes it a promising candidate for applications like PVs, where strong light absorption is essential. Ca$_3$PBr$_3$ and Ca$_3$PCl$_3$ show absorption characteristics similar to Ca$_3$PI$_3$ but with slightly less intensity. Ca$_3$PF$_3$ displays a weaker absorption, especially in the lower energy range, making it less suitable for light harvesting applications.

%%%%%%%%%%%%%%%%%%%%%%%fig%%%%%%%%%%%%%%%%%%%%%%%%%%

\subsubsection{Refractive index and extinction coefficient}
\begin{figure}[h]
	\centerline{\includegraphics[scale=0.3]{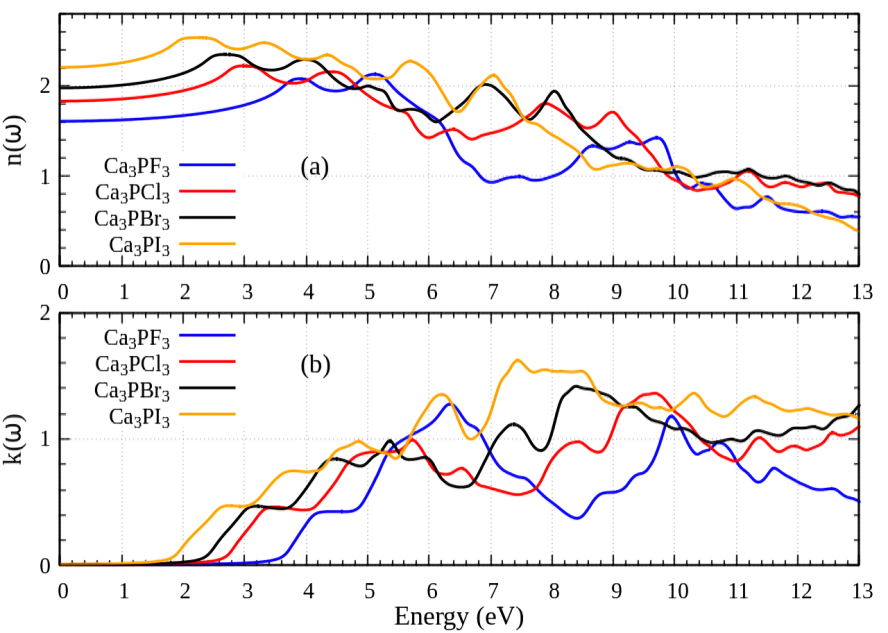}}
	\caption{(Colour online) Variation in (a) $n(\omega)$ and (b) $k(\omega)$  for Ca$_3$PX$_3$ (X = F, Cl, Br, I) compounds.} \label{fig:7}
\end{figure}

Figures~\ref{fig:7}(a) and \ref{fig:7}(b) show the refractive index, $n(\omega)$, and the extinction coefficient, $k(\omega)$, respectively. In the $n(\omega)$ plot, all materials show an increasing refractive index with energy, with notable peaks around the 2 eV to 4 eV range. Ca$_3$PI$_3$ shows the highest refractive index at lower energies, indicating a stronger interaction between light and the material in this energy range. 
%%%%%%%%%%%%%%%%%%%%%%%FIG%%%%%%%%%%%%%%%%%%%%%%%%%%%%%%%%%%%%
%%%%%%%%%%%%%%%%%%%%%%%FIG%%%%%%%%%%%%%%%%%%%%%%%
The refractive index is important for understanding how light bends when passing through the material, and a higher refractive index suggests that the material can guide light more effectively, which is crucial for optical devices like waveguides and light-coupling systems. The static refractive index values for the compounds studied varied from 1.61 (for Ca$_3$PF$_3$) to 2.20 (for Ca$_3$PI$_3$) progressively increasing with the energy of the incident EM radiation in the infrared and visible regions and then eventually decreasing being still positive while energy increased as the materials remained transparent. This confirmed the isotropic nature of the materials by which their refractive index did not depend on the direction. The extinction coefficient, $k$, greatly influenced the capability of a substance to absorb light such that a high $k$ meant a more efficient absorption of the EM radiation. The extinction coefficient varied after reaching  its maximum value when the energy of the incident radiation was varied, as displayed in figure~\ref{fig:7}(b). For the investigated materials, it reached to peak values in the range 6--8 eV. Ca$_3$PI$_3$ exhibits the highest extinction coefficient, suggesting that it absorbs light most strongly in this region, making it an ideal candidate for applications that require high light absorption, such as solar cells.

\subsubsection{Reflectivity and energy loss function}

\begin{figure}[htb]
	\centerline{\includegraphics[scale=0.3]{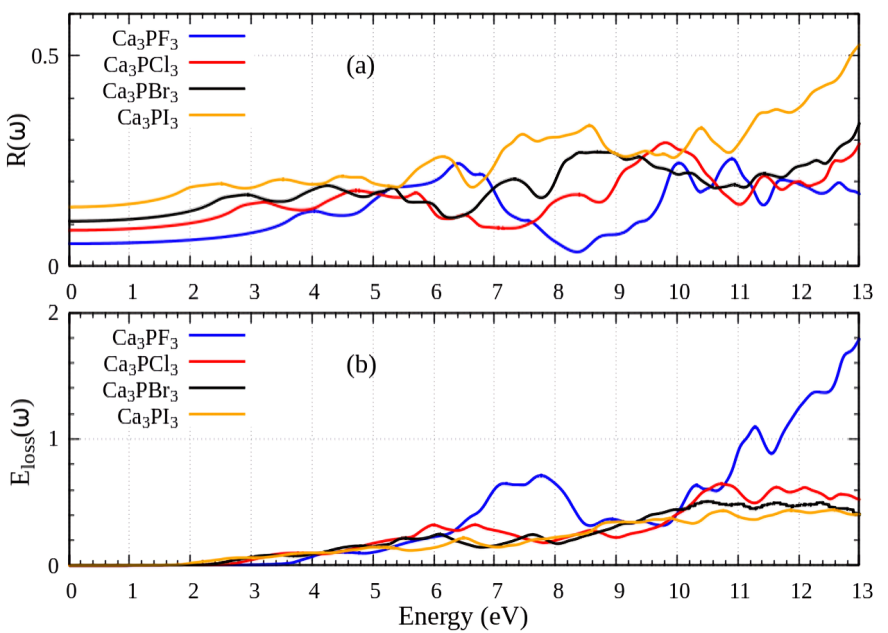}}
	\caption{(Colour online) Variation in (a) $R(\omega)$ and (b) $E_{\rm loss}(\omega)$ for Ca$_3$PX$_3$ (X = F, Cl, Br, I) compounds.} \label{fig:8}
\end{figure}

Reflectivity determines the fraction of incident light reflected by a given substance. This graph shows how reflectivity as a function of photon energy varies in the range specified. In part (a) of figure~\ref{fig:8} the optical reflectivity $R(\omega)$ is presented, and it can be seen that the static reflectivity values, $R(0)$, of Ca$_3$PX$_3$ (X = F, Cl, Br, I) are listed as 5\%, 9\%, 11\%, and 14\%, respectively. Throughout the entire range of incident radiation, $R(\omega)$ generally increases with energy for all the materials, especially above~2.5~eV. This increase is more pronounced in Ca$_3$PI$_3$, which shows a sharp rise, especially in the higher energy range (above 6 eV), indicating that the material is highly reflective at these energies. This is important for applications that require high reflectance, such as in optical coatings or mirrors in photonic devices. Ca$_3$PF$_3$, by contrast, exhibits the lowest reflectivity, suggesting less interaction with light at higher energies, which could limit its use in applications where high reflectance is critical. The energy decay of a moving electron within a substance can be characterized by its energy loss function, $E_{\rm loss}(\omega)$. It is crucial in applications like plasmonics and solar cells, where materials with higher energy loss can contribute to an efficient light-to-heat conversion. Figure~\ref{fig:8}(b) demonstrates the change in $E_{\rm loss}(\omega)$ as the energy of incident radiation varies. A notable feature across all materials is a more pronounced increase in energy loss above 5 eV, where the materials begin to absorb energy more efficiently. This suggests the onset of electronic transitions or excitonic effects that result in greater energy dissipation. Table~\ref{tab:2} presents the computed static values of the dielectric constant $\varepsilon_{0}$, refractive index $n(0)$, and reflectivity $R(0)$ along with a comparison to other available values.

\begin{table}[htb]
	\caption{Optical parameters of Ca$_3$PX$_3$ (X = F, Cl, Br, I) materials.}
	\label{tab:2}
	%\vspace{2ex}
	\begin{center}
		\renewcommand{\arraystretch}{0}
		\begin{tabular}{|c|c|c|c|c|}
			\hline
			\textbf{Compounds}&\textbf{Dielectric Constant}&\textbf{Refractive Index}&\textbf{Reflectivity}&\textbf{SLME}\strut\\
			&$\varepsilon_{1}(0)$&$n(0)$&$R(0)$&\%\strut\\
			\hline
			Ca$_3$PF$_3$&2.58&1.61&0.05&0.60\strut\\
			\hline
			Ca$_3$PCl$_3$&3.34&1.83&0.09&8.30\strut\\
			&5.97$^{a}$&&&\strut\\
			\hline
			Ca$_3$PBr$_3$&3.90&1.97&0.11&14.75\strut\\
			&5.36$^{a}$&&&\strut\\
			\hline
			Ca$_3$PI$_3$&4.85&2.20&0.14&29.60\strut\\
			&5.91$^{a}$&&&\strut\\
			\hline
			\multicolumn{5}{l}{$^{a}$Reference~\cite{ref16}} \\
		\end{tabular}
		\renewcommand{\arraystretch}{1}
	\end{center}
\end{table}

\subsection{Analysis of spectroscopic limited maximum efficiency (SLME)} 

\begin{figure}[htb]
	\centerline{\includegraphics[scale=0.3]{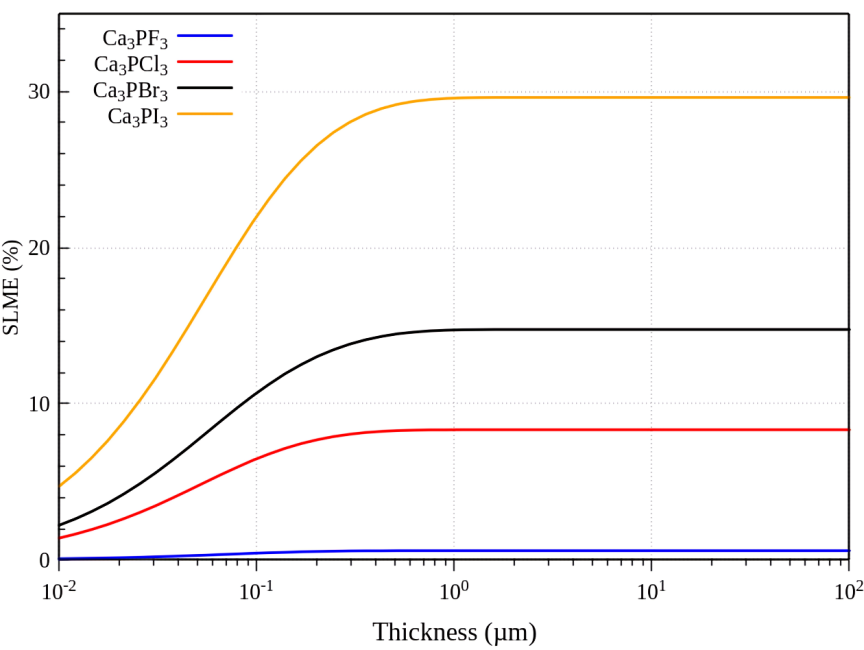}}
	\caption{(Colour online) SLME curves for Ca$_3$PX$_3$ (X = F, Cl, Br, I) compounds.} \label{fig:9}
\end{figure}

The efficiency of PV devices is influenced by several factors, including the fabrication method of the device, optical properties of the materials, and presence of defects within the material. To identify suitable materials for PV applications, Yu and Zunger~\cite{ref26} have introduced the concept of SLME as an effective screening technique. The SLME is calculated by considering the absorption spectrum of the material along with the standard AM 1.5G solar spectrum at 25$^\circ$C \cite{ref33}. The SLME for the Ca$_3$PX$_3$ (X = F, Cl, Br, I) materials, as a function of film thickness, was calculated using these spectra. The variation in the SLME with respect to the film thickness is shown in figure~\ref{fig:9}. The SLME graphs show a similar trend of increasing efficiency with thickness, followed by a plateau after reaching a certain value. The maximum SLME for each material varies, reflecting differences in their light absorption properties and band structures, and is tabulated in table~\ref{tab:2}. 

%%%%%%%%%%%%%%%%%%%%%%%%SLMEFIG%%%%%%%%%%%%%%%%%%%%%%

Among these materials, Ca$_3$PI$_3$ exhibits the highest SLME, reaching around 30\%, with a minimum film thickness requirement of 1  \textmu{}m. This material shows a  better performance due to its enhanced light absorption properties. The SLME shows an increasing trend with film thickness, indicating that thicker films allow for greater absorption of light and, thus, higher efficiency. However, despite this improvement, the SLME remains somewhat lower compared to the Shockley--Queisser (SQ) limit for single-junction solar cells, which is around 33\%~\cite{ref34}, suggesting that its band structure or absorption characteristics may limit its performance. Ca$_3$PBr$_3$ reaches a maximum SLME of around 15\% and appears to hold a good potential for use in high-efficiency single-junction solar cells and as a part of a multi-junction solar cell. In comparison to Ca$_3$PBr$_3$, Ca$_3$PCl$_3$ seems to have slightly worse performance with a maximum SLME of around 8\%. 

The higher SLME for Ca$_3$PBr$_3$ compared to Ca$_3$PCl$_3$ suggests a superior light absorption; with both materials producing SLME lower than the SQ limit. The performance of both Ca$_3$PBr$_3$ and Ca$_3$PCl$_3$ appear favourable for use in multi-junction devices, as the low efficiencies of the materials can be enhanced by using other materials with higher efficiency potential. On the other hand, Ca$_3$PF$_3$ reaches a maximum SLME of around 0.6\%, which is the lowest amongst all materials investigated, indicating that it suffers from either poor light absorption or that the large direct bandgap restricts the photon absorption which adversely affects the performance. Therefore, although low efficiency of Ca$_3$PF$_3$ makes it poorly suited to high-performance applications, it can potentially serve for certain applications where lower efficiency is acceptable.

\section{Summary and conclusions}
%\fcolorbox{yellow}{yellow}{
This study presents a first-principle analysis of the structural, electronic and optical properties of the Ca$_3$PX$_3$ (X = F, Cl, Br, I) halide compounds. All of these materials show structural stability in cubic phase Pm$\bar{3}$m, with lattice constants in the range 5.321 \AA~(Ca$_3$PF$_3$) to 6.20 \AA~(Ca$_3$PI$_3$). The calculated electronic structure computed using the PBE+TB-mBJ method shows a  direct bandgap of these compounds in the range of 3.788 eV (Ca$_3$PF$_3$) to 2.0 eV (Ca$_3$PI$_3$), demonstrating the increasing degree of optical activity towards the heavier members of this series of halides. The optical constants of each compound also increase progressively; while the optical response of Ca$_3$PF$_3$ has a static dielectric constant of 2.58 and 4.85 for Ca$_3$PI$_3$, the refractive index ranges from 1.61 to 2.20 for the members of this series. In particular, the optical properties of Ca$_3$PI$_3$ show the greatest potential for converting energy into an electrical form by exhibiting a significant absorption in the visible range, as well as exhibiting the highest photovoltaic potential. The SLME values further support this trend; guiding Ca$_3$PI$_3$ and Ca$_3$PBr$_3$ as the most efficient candidates. Therefore, among all the  materials studied here, Ca$_3$PI$_3$ evolved as the most promising material for future optoelectronic and photovoltaic devices. Future studies may investigate the effect of spin-orbit coupling, excitonic corrections, as well as transport through defects to validate these materials for practical solar-cell integration.

%\subsection*{Declaration}
%\textbf{Ethical Approval:} Not applicable in this work.\\

%\textbf{Conflicat of interst:} This manuscript does not include a conflict of interest.\\

%\textbf{Funding:} No funding was received for this work.\\

%\textbf{Availability of data:} Data and materials will be available on reasonable request.  

\subsection*{Acknowledgement}
First author Priyanka Dhariwal thankful to CSIR, New Delhi, India for providing financial assistance in form of Junior Research Fellowship (JRF).

\ukrainianpart

\title{Вплив галогенідного заміщення на фотоелектричні властивості перовскітів Ca$_3$PX$_3$ (X = F, Cl, Br, I): підвищення ефективності сонячних елементів}
\author{П. Дхарівал\refaddr{label1},
	Д. Пракаш\refaddr{label2}, 
	К. Д. Верма\refaddr{label1}, 
	А. Кумарі\refaddr{label1}, 
	П. К. Камлеш\refaddr{label3}, 
	А.~С.~Верма\refaddr{label4,label5}}
\addresses{
	\addr{label1} Науково-дослідна лабораторія матеріалознавства, фізичний факультет, коледж Шрі Варшні, університет ім.~Раджі Махендри Пратапа Сінгха, Алігарх, 202001 Уттар-Прадеш, Індія
	\addr{label2} Школа комп’ютерних наук та інженерії, інженерний факультет, Університет Шрі Мата Вайшно Деві, Какр'ял, Катра, 182320 J\&K, Індія
	\addr{label3} Фізичний факультет університету Пурніма, Джайпур, 303905 Раджастан, Індія
	\addr{label4} Кафедра суміжних наук, Університет графічної ери, Дехрадун, 248002 Уттаракханд, Індія
	\addr{label5} Університетський центр досліджень та розробок, кафедра фізики, Чандігархський університет, 140413 Мохалі, Пенджаб, Індія
}

\makeukrtitle

\begin{abstract}
	\tolerance=3000%
	Досліджено фундаментальні фізичні характеристики такі як структурні, електронні та оптичні параметри матеріалів Ca$_3$PX$_3$ (X = F, Cl, Br, I) з точки зору їх потенційного застосування в оптоелектроніці, зокрема для сонячних елементів та пов'язаних з ними пристроїв. Згідно з кристалографічними дослідженнями, Ca$_3$PI$_3$ має найстабільнішу конфігурацію серед усіх досліджених матеріалів. Аналіз зонної структури цих матеріалів показує, що вони мають пряму заборонену зону в діапазоні від 2,0~еВ до 3,788~еВ, що робить їх ідеальними для поглинання світла. Для фотоелектричних застосувань, ми проаналізували обмежену максимальну ефективність методом спектроскопічного скринінгу з перших принципів, яка підтверджує, що матеріал Ca$_3$PI$_3$ демонструє найвищу ефективність сонячних елементів 29,6\%, а Ca$_3$PF$_3$ показує нижчу ефективність 0,6\% для придатності сонячних елементів. Таким чином, отримані результати підтверджують реальний потенціал та можливості галогенідного заміщення для налаштування матеріалів щодо використання у певних оптоелектронних пристроях.
	\keywords ефекти заміщення, оптоелектронні застосування, показник заломлення, діелектричні властивості
	
\end{abstract}

\lastpage
\end{document}